\documentclass[superscriptaddress,amsmath,amssymb,aps,prb,twocolumn,showkeys,nofootinbib]{revtex4-2}
\usepackage{graphicx}         
\usepackage{dcolumn}          
\usepackage{bm}               
\usepackage{upgreek}
\usepackage{booktabs} 				
\usepackage{tabularx}
\usepackage{array}
\usepackage[colorlinks=true,urlcolor=blue,breaklinks=true]{hyperref}

\usepackage{xcolor}

\usepackage{xr}

\usepackage{siunitx}
\usepackage{stackengine}
\usepackage{float}

\newsavebox{\twosubbox}

\begin{document}

\title{Aharonov-Bohm interference and phase-coherent surface-state transport in topological insulator rings}

\author{Gerrit Behner}
\email{g.behner@fz-juelich.de}
\affiliation{Peter Gr\"unberg Institut (PGI-9), Forschungszentrum J\"ulich, 52425 J\"ulich, Germany}
\affiliation{JARA-Fundamentals of Future Information Technology, J\"ulich-Aachen Research Alliance, Forschungszentrum J\"ulich and RWTH Aachen University, Germany}

\author{Abdur Rehman Jalil}
\affiliation{Peter Gr\"unberg Institut (PGI-9), Forschungszentrum J\"ulich, 52425 J\"ulich, Germany}
\affiliation{JARA-Fundamentals of Future Information Technology, J\"ulich-Aachen Research Alliance, Forschungszentrum J\"ulich and RWTH Aachen University, Germany}

\author{Dennis Heffels}
\affiliation{Peter Gr\"unberg Institut (PGI-9), Forschungszentrum J\"ulich, 52425 J\"ulich, Germany}
\affiliation{JARA-Fundamentals of Future Information Technology, J\"ulich-Aachen Research Alliance, Forschungszentrum J\"ulich and RWTH Aachen University, Germany}

\author{Jonas Kölzer}
\affiliation{Peter Gr\"unberg Institut (PGI-9), Forschungszentrum J\"ulich, 52425 J\"ulich, Germany}
\affiliation{JARA-Fundamentals of Future Information Technology, J\"ulich-Aachen Research Alliance, Forschungszentrum J\"ulich and RWTH Aachen University, Germany}

\author{Kristof Moors}
\affiliation{Peter Gr\"unberg Institut (PGI-9), Forschungszentrum J\"ulich, 52425 J\"ulich, Germany}
\affiliation{JARA-Fundamentals of Future Information Technology, J\"ulich-Aachen Research Alliance, Forschungszentrum J\"ulich and RWTH Aachen University, Germany}

\author{Jonas Mertens}
\affiliation{Peter Gr\"unberg Institut (PGI-9), Forschungszentrum J\"ulich, 52425 J\"ulich, Germany}
\affiliation{JARA-Fundamentals of Future Information Technology, J\"ulich-Aachen Research Alliance, Forschungszentrum J\"ulich and RWTH Aachen University, Germany}

\author{Erik Zimmermann}
\affiliation{Peter Gr\"unberg Institut (PGI-9), Forschungszentrum J\"ulich, 52425 J\"ulich, Germany}
\affiliation{JARA-Fundamentals of Future Information Technology, J\"ulich-Aachen Research Alliance, Forschungszentrum J\"ulich and RWTH Aachen University, Germany}

\author{Gregor Mussler}
\affiliation{Peter Gr\"unberg Institut (PGI-9), Forschungszentrum J\"ulich, 52425 J\"ulich, Germany}
\affiliation{JARA-Fundamentals of Future Information Technology, J\"ulich-Aachen Research Alliance, Forschungszentrum J\"ulich and RWTH Aachen University, Germany}

\author{Peter Schüffelgen}
\affiliation{Peter Gr\"unberg Institut (PGI-9), Forschungszentrum J\"ulich, 52425 J\"ulich, Germany}
\affiliation{JARA-Fundamentals of Future Information Technology, J\"ulich-Aachen Research Alliance, Forschungszentrum J\"ulich and RWTH Aachen University, Germany}

\author{Hans L\"uth}
\affiliation{Peter Gr\"unberg Institut (PGI-9), Forschungszentrum J\"ulich, 52425 J\"ulich, Germany}
\affiliation{JARA-Fundamentals of Future Information Technology, J\"ulich-Aachen Research Alliance, Forschungszentrum J\"ulich and RWTH Aachen University, Germany}

\author{Detlev Gr\"utzmacher}
\affiliation{Peter Gr\"unberg Institut (PGI-9), Forschungszentrum J\"ulich, 52425 J\"ulich, Germany}
\affiliation{JARA-Fundamentals of Future Information Technology, J\"ulich-Aachen Research Alliance, Forschungszentrum J\"ulich and RWTH Aachen University, Germany}

\author{Thomas Sch\"apers}
\email{th.schaepers@fz-juelich.de}
\affiliation{Peter Gr\"unberg Institut (PGI-9), Forschungszentrum J\"ulich, 52425 J\"ulich, Germany}
\affiliation{JARA-Fundamentals of Future Information Technology, J\"ulich-Aachen Research Alliance, Forschungszentrum J\"ulich and RWTH Aachen University, Germany}

\hyphenation{}
\date{\today}


\begin{abstract}
We present low-temperature magnetotransport measurements on selectively-grown Sb$_2$Te$_3$-based topological insulator ring structures. These topological insulator  ring geometries display clear Aharonov-Bohm  oscillations in the conductance originating from phase-coherent transport around the ring. The temperature dependence of the oscillation amplitude indicates that the Aharonov-Bohm oscillations originate from ballistic transport along the ring arms. The oscillations can therefore be attributed to topological surface states, which can maintain a quasi-ballistic transport regime in the presence of disorder. Further insight on the phase coherence is gained by comparing with similar Aharonov-Bohm-type oscillations in topological insulator  nanoribbons exposed to an axial magnetic field. Here, quasi-ballistic phase-coherent transport is confirmed for closed-loop topological surface states in transverse direction enclosing the cross-section of the nanoribbon. In contrast, the appearance of universal conductance fluctuations indicates phase-coherent transport in the diffuse regime, which is attributed to bulk carrier transport. Thus, it appears that even in the presence of diffusive $p$-type charge carriers in Aharonov-Bohm ring structures, phase-coherent quasi-ballistic transport of topologically protected surface states is maintained over long distances.
\end{abstract}

\maketitle

\section{Introduction}

Phase-coherence has a great impact on transport in mesoscopic systems which leads to many interesting effects, visible through their quantum mechanical correction to the conduction as a function of magnetic field or gate voltage~\cite{Beenakker91,Lin02}. Typical phenomena associated with phase-coherent transport are weak (anti-)localization, universal conductance fluctuations (UCFs), or Aharonov-Bohm (AB) oscillations~\cite{Beenakker91}. Recently, phase-coherent transport has also been studied in three-dimensional topological insulators (TIs) such as Bi$_2$Te$_3$, Sb$_2$Te$_3$, Bi$_2$Se$_3$, or their alloys, in which topologically protected spin-momentum locked surface states are present~\cite{Hasan10,Qi11}. Interest in these materials stems from applications in topoelectronic circuits and topological quantum computer architectures~\cite{Nayak08,Hyart13,Sarma15,Aasen16}.

\begin{figure}
    \centering
    \includegraphics[width=0.75\linewidth]{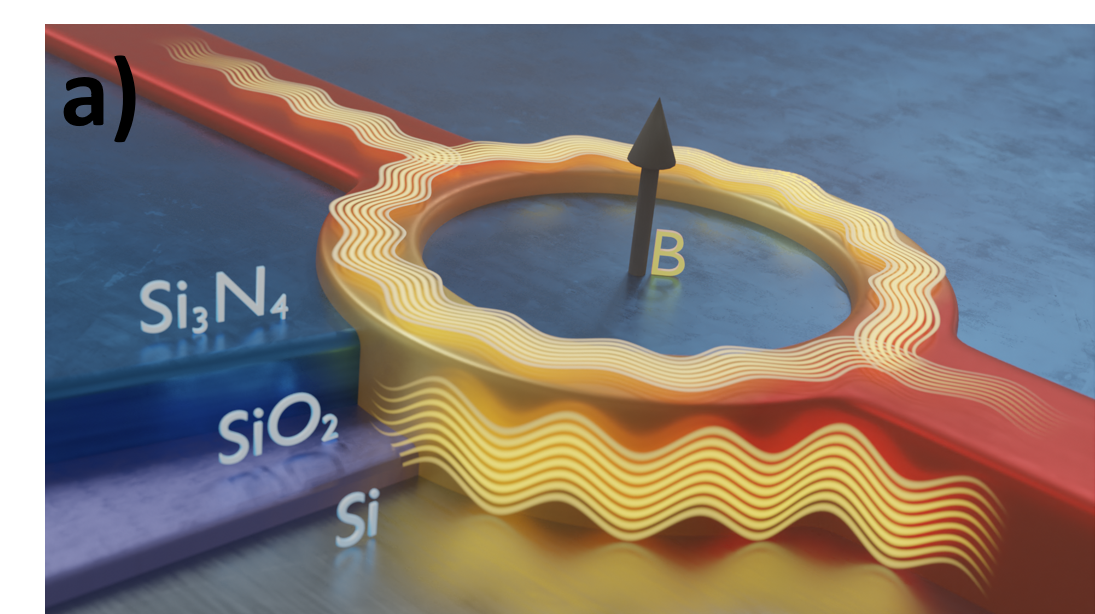}
    \includegraphics[width=0.75\linewidth]{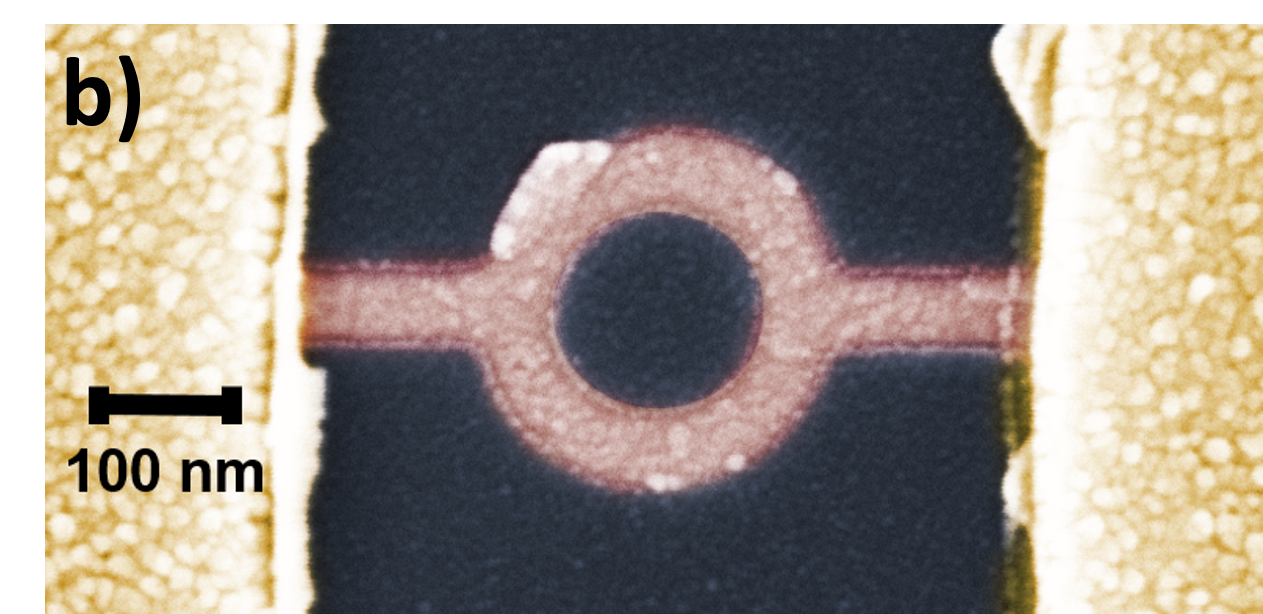}
    \caption{
    Aharonov-Bohm interferometer: (a) Schematics of the selectively-grown ring structure. (b) Scanning electron micrograph of a selectively-grown Sb$_2$Te$_3$-based ring device (sample A) with an inner and outer radius of 100\,nm and 150\,nm, respectively.
    } 
    \label{fig:Device}
\end{figure}

In previous studies, various transport properties of straight three-dimensional TI-based nanowires and nanoribbons have been investigated theoretically and studied experimentally in micrometer- and nanometer-sized systems. The observed effects range from weak antilocalization and conductance fluctuations to the manifestation of quasi-ballistic transport of topologically protected surface states, inducing Aharonov-Bohm-type conductance oscillations when applying a magnetic field along the wire or ribbon~\cite{Peng10,Xiu11,Dufouleur13,Arango16,Jauregui15,Koelzer20,Rosenbach20,Kim20}.

Three-dimensional TIs often tend to be intrinsically doped due to the formation of crystal defects during growth, leading to an additional bulk transport channel as the Fermi level either crosses the conduction or valence band~\cite{Lostak89,Scanlon12}. With respect to phase-coherent transport, it is a difficult task to disentangle the contributions from bulk and topologically protected surface states. On the one hand, the picture regarding UCFs is not so clear. In nanoribbons of ternary materials with a relatively small bulk contribution, the UCFs have been attributed to surface states~\cite{Koelzer20}, while in Bi$_2$Se$_3$ nanoribbons they have been assigned to bulk carriers~\cite{Dufouleur16}. On the other hand, the AB-type oscillations observed in nanoribbons are generally believed to be due to phase-coherent loops formed by topologically protected surface states in the transverse direction around the perimeter of the cross section~\cite{Peng10,Xiu11,Dufouleur13,Arango16,Jauregui15,Dufouleur18,Rosenbach20,Kim20}. From the exponential decrease of the oscillation amplitude with temperature, it was deduced that the transport is quasi-ballistic~\cite{Dufouleur13,Rosenbach20,Kim20}. However, not much information has been available on phase-coherent transport of topologically protected surface states along the axial direction of the nanoribbon, i.e., the direction along which the (surface-state) current flows. To address this issue, we measure planar Sb$_2$Te$_3$ ring-shaped interferometers and investigate the Aharonov-Bohm effect with an out-of-plane magnetic field. Here, we identify a clear peak in the Fourier spectrum of the magnetoconductance corresponding to magnetic flux quantum-periodic oscillations. From the decrease of that peak with temperature, the corresponding transport regime is identified. The comparison of different ring sizes as well as a detailed analysis of the spectrum of the UCFs and the Aharonov-Bohm oscillations in straight nanoribbons allow a comprehensive investigation of the phase-coherence in the different transport channels. The interpretation of the transport measurement data is supported by quantum transport simulations.  

\section{Experimental}

The Sb$_2$Te$_3$ layer was grown by molecular beam epitaxy employing a selective-area growth approach \cite{Kampmeier16}. For the substrate preparation, first, 5\,nm of a Si(111) wafer was thermally converted into SiO$_2$. Subsequently, a 20-nm-thick Si$_3$N$_4$ layer was deposited by low-pressure chemical vapor deposition. The ring and nanoribbon structures were defined by electron beam lithography. Using reactive ion etching (CHF$_3$/O$_2$) and hydrofluoric acid wet etching, the Si$_3$N$_4$ and the SiO$_2$ layers were etched, respectively, to reveal the Si(111) surface locally. The structured Si$_3$N$_4$/SiO$_2$ layers formed the selective-area growth mask. The standard parameters for selective growth of Sb$_2$Te$_3$, given a substrate temperature of $300\,^\circ$C, a Sb-cell temperature of $470^\circ$C, and a Te-cell temperature of $325\,^\circ$C, resulted in a growth rate of 7\,nm/h. The 20-nm-thick TI ring and nanoribbon structures were grown in the Te-overpressure regime. To prevent oxidation the Sb$_2$Te$_3$ layer was capped by a 5-nm-thick AlO$_x$ layer. From Hall measurements at \SI{1.5}{K} we determined a hole carrier concentration of $7.4 \times 10^{13} \, \textnormal{cm}^{-2}$ and a mobility of \SI{152}{cm^{2}/Vs} (see Supplementary Material~SI).

The Ohmic contacts composed of a 5-nm-thick Nb layer and a 100-nm-thick Au layer were deposited on top of the TI layer after removing the AlO$_x$ capping in the contact areas by wet chemical etching and argon sputtering. Rings of two different sizes were investigated, i.e., samples A and B with an outer radius of \SI{150}{nm} and \SI{200}{nm}, respectively, both with an annulus width of \SI{50}{nm}. Figure~\ref{fig:Device} shows a schematic of a selectively-grown ring structure as well as a scanning electron micrograph of sample A. To further characterize the properties of the Sb$_2$Te$_3$ layers, a 100-nm-wide nanoribbon (sample C) prepared in the same run as the ring structures was fabricated. The basic transport properties were extracted from a 500-nm-wide Hall bar structure (see Supplementary Material SI).

The measurement of the ring structures were conducted in a $^3$He-cryostat with a base temperature of \SI{400}{mK}. The magnetic field is applied perpendicularly to the substrate plane using a superconducting magnet. The magnetotransport of the nanoribbon and Hall bar were measured in a variable temperature insert with a base temperature of \SI{1.4}{K}. The current was measured using a standard two-probe voltage sensing lock-in setup for all samples.

\section{Results and discussion}

\subsection{Aharonov-Bohm effect in ring structures}

\begin{figure*}
    \centering
    \includegraphics[width=0.95\linewidth]{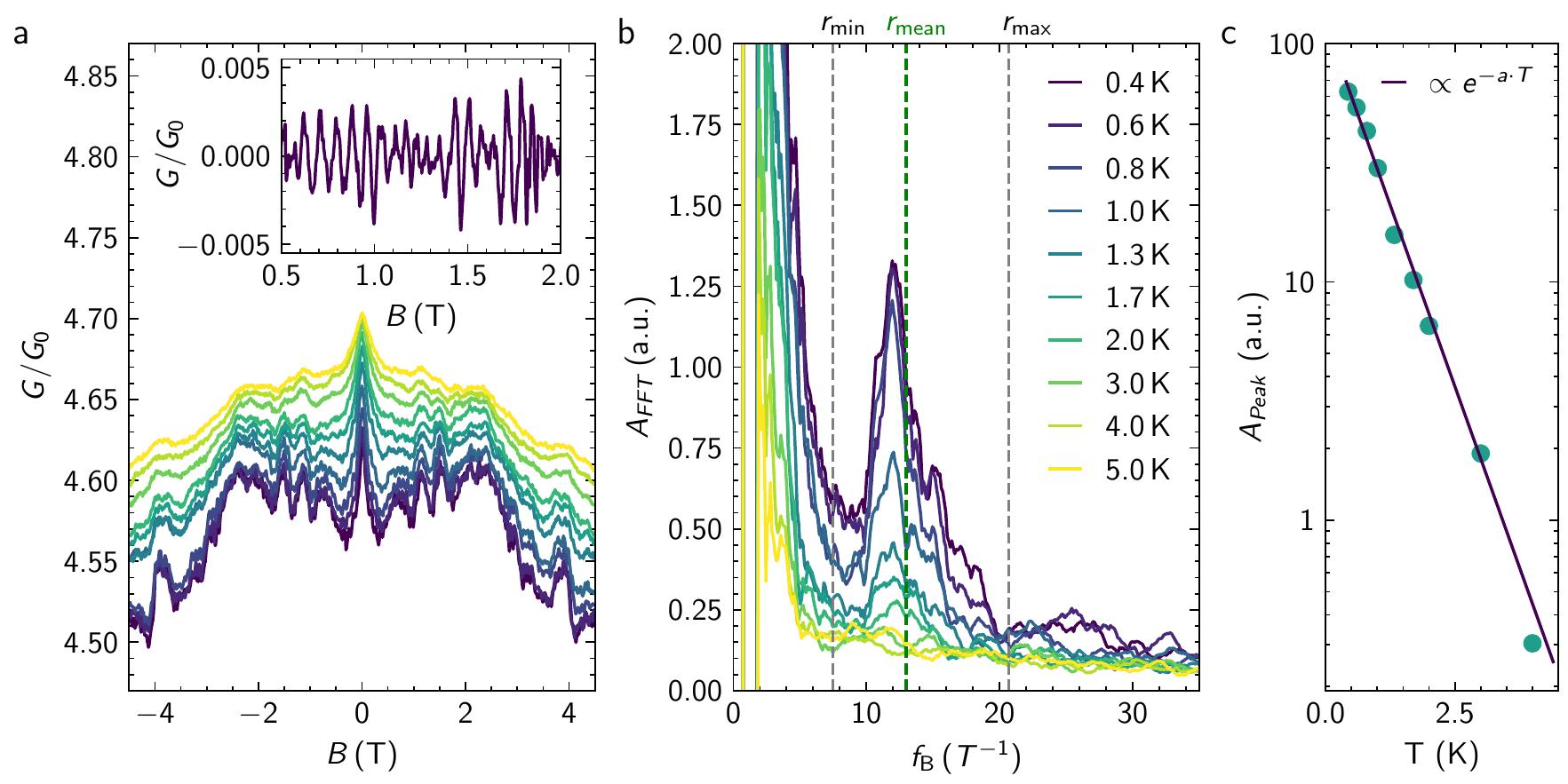}
    \caption{Magnetotransport of ring structures:
    (a) Normalized magnetoconductance of sample A at temperatures in the range of \SI{0.4}{K} to \SI{5.0}{K}, with $G_0=2e^2/h$. The inset shows a detail of the oscillations at \SI{800}{mK} in a smaller magnetic field range. The measured period $\Delta B$ of \SI{82.5}{mT}, fits accurately to the peak observed in the FFT of the measured data. The legend for the different temperatures is given in (b).
    (b) Fourier spectrum of the magnetoconductance shown in (a). The expected frequency according to the minimum and maximum radius as well as the mean radius are indicated by dashed lines.
    (c) Integrated amplitude $\mathcal{A}$ of the peak in the FFT at \SI{12}{T^{-1}} as a function of temperature. The solid line represents an exponential fit.} 
    \label{fig:Sb2Te3-100nm-ring-AB}
\end{figure*}

Figure~\ref{fig:Sb2Te3-100nm-ring-AB}a shows the normalized magnetoconductance $G/G_0$ of a ring interferometer structure (sample A) with $G_0=2e^2/h$. The measurement temperature is varied from \SI{0.4}{K} to \SI{5.0}{K}. The magnetic field is oriented perpendicularly to the substrate plane so that a magnetic flux is threading the ring aperture. The corresponding data of the second ring structure (sample B) is presented in the Supplementary Material SII.

The magnetoconductance exhibits several features. The most striking one is a peak at zero magnetic field, which is due to the weak antilocalization effect. This peak structure has been observed previously in TI nanoribbon structures of similar width and is due to electron interference combined with strong spin-orbit coupling (see Supplementary Material SIII)~\cite{Koelzer20,Rosenbach20}. Another feature are pronounced conductance fluctuations with larger amplitude over larger B-field intervals, in particular at low temperatures. These are caused by the interference of a limited number of trajectories due to the small dimensions of the sample~\cite{Lee87}, which will be discussed in more detail at a later stage to provide complementary information on phase-coherence. A closer look at the magnetoconductance reveals that regular oscillations with a smaller amplitude are superimposed on the conductance fluctuations. A magnification of a smaller magnetic field region is shown in the inset of Fig.~\ref{fig:Sb2Te3-100nm-ring-AB}a. We found that the oscillation period is about $\Delta B=\SI{82.5}{mT}$. We attribute these regular features to the Aharonov-Bohm effect in the ring-shaped conductor~\cite{Aharonov59}. Indeed, the period $\Delta B$ fits very well to oscillations with a flux period of $\phi_0=\Delta B \cdot A$, with $\phi_0=h/e$ the magnetic flux quantum and $A=\pi r_\mathrm{mean}^2$ the area of the disc with radius equal to the mean radius of the ring, $r_\mathrm{mean}$ of \SI{125}{nm}.

The periodic features in the magnetoconductance are analyzed using a fast Fourier transform (FFT), as shown in Fig.~\ref{fig:Sb2Te3-100nm-ring-AB}b. Note that the Fourier transform was applied to the original data without any filtering. The FFT shows a distinct peak at a frequency $f_B$ of \SI{12}{T^{-1}}, which corresponds to the expected value for the mean radius of the ring indicated by the vertical dashed green line in Fig.~\ref{fig:Sb2Te3-100nm-ring-AB}b. In general, the peak lies within the frequency limits given by the inner ($r_\mathrm{min}$) and outer ($r_\mathrm{max}$) ring radii, indicating that the trajectories of the electron partial waves cover the entire ring area. As the temperature increases, the height of the peak decreases corresponding to a reduction of the oscillation amplitude. At about \SI{3.0}{K}, the peak has disappeared. In addition to the peak at about \SI{12}{T^{-1}}, a weaker feature is observed at about \SI{25}{T^{-1}}, where the second harmonic is expected.

From the decrease of the integrated peak height $\mathcal{A}(T)$ at \SI{12}{T^{-1}} in the FFT with increasing temperature, we estimate the phase-coherence length $l_\varphi$. The integration is performed within a window bounded by the frequencies corresponding to flux quantum periodicity when considering the inner and outer radius of the ring (cf. Fig.~\ref{fig:Sb2Te3-100nm-ring-AB}b). We consider an exponential decay $\mathcal{A} \propto \exp(-\pi r_\mathrm{mean}/l_\varphi(T))$~\cite{Washburn86}, with $\pi r_\mathrm{mean}= \SI{393}{nm}$ the length of one of the ring arms and $\mathcal{A}$ being a measure of the oscillation amplitude. Indeed, the peak height decay is very well fitted by an exponential decrease with $l_\varphi(T) \sim T^{-1}$, as shown in Fig.~\ref{fig:Sb2Te3-100nm-ring-AB}c. From the fit, we obtain a phase-coherence length of $l_\varphi = \SI{722}{nm}$ at a temperature of \SI{0.4}{K}, which is considerably longer than the length of the ring arm.

\begin{figure*}
    \centering
    \includegraphics[width=0.95\linewidth]{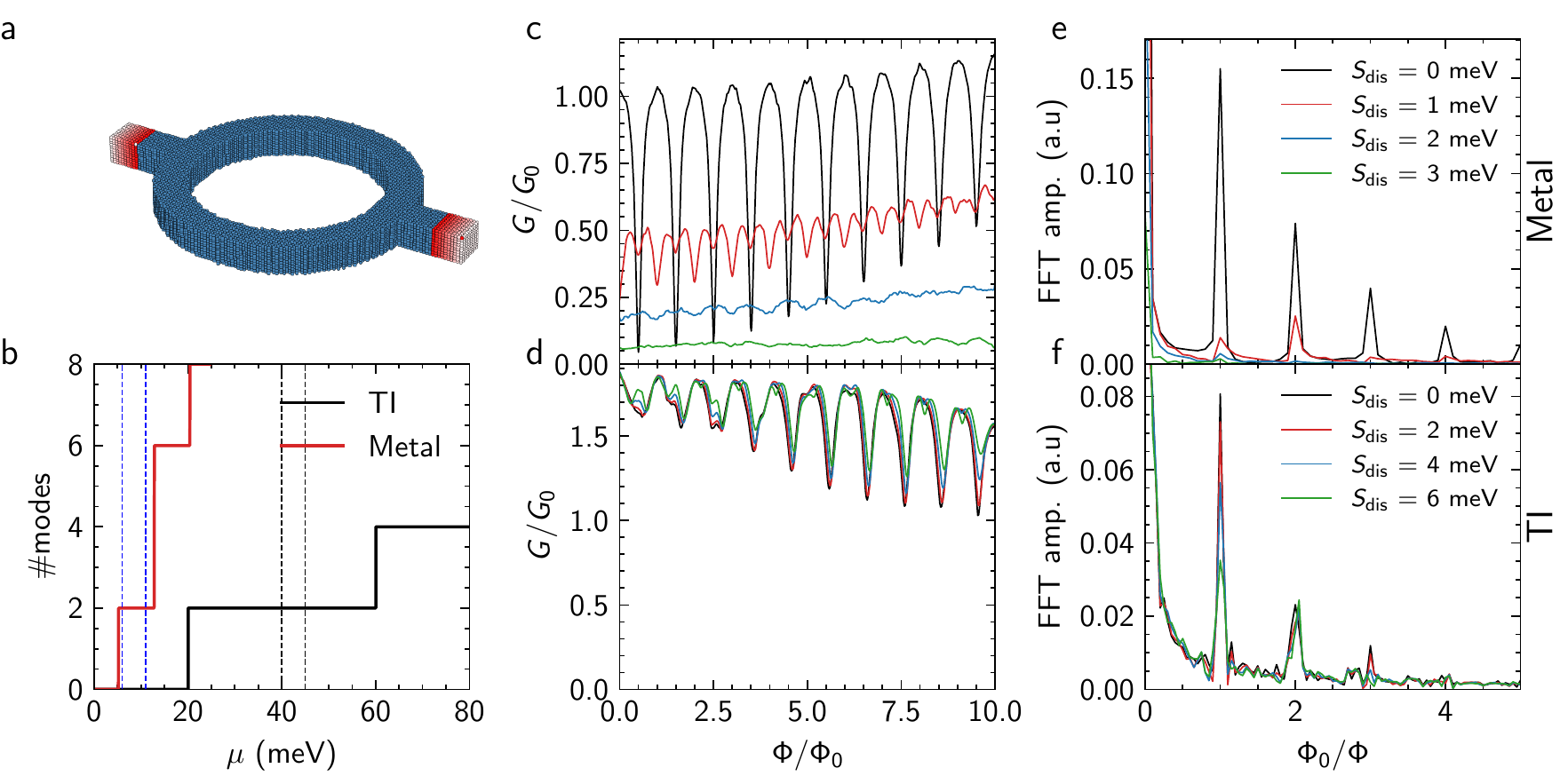}
    \caption{Tight-binding simulations of a metallic and topological insulator ring structure:
    (a) Schematic of the tight-binding model, with semi-infinite leads (a few unit cells indicated in red) attached to a (disordered) scattering region (in blue).
    (b) The number of transport channels as a function of energy for a bulk metallic ring and a (bulk-insulating) TI ring. The vertical dashed lines delineate the energy window over which the magnetoconductance is averaged.
    (c),(d) The conductance as a function of the magnetic flux threading the aperture of (c) a bulk metallic ring and (d) a (bulk-insulating) TI ring for different values of the disorder strength in the scattering region. (e),(f) The Fourier transform of the conductance profile in (c) and (d), respectively.}
    \label{fig:Conductance}
\end{figure*}

Aharonov-Bohm oscillations arise when the phase-coherence length $l_\varphi$ is of the order of the length of a ring arm, which is the case for our ring, as shown above. In principle, in our intrisically doped TI samples, the oscillations can originate from bulk carriers as well as from charge carriers in topologically protected surface states. However, the exponential decay of the FFT amplitude with $l_\varphi(T) \sim T^{-1}$ indicates that the transport is in the quasi-ballistic mesoscopic regime~\cite{Seelig01}. Hence, we anticipate that the observed AB oscillations are mainly due to transport of topologically protected surface states. Indeed, in TI nanoribbons it was deduced that these surface states with spin-momentum locking have an enhanced transport mean free path due to strong anisotropic scattering \cite{Culcer10,Dufouleur16}. This is consistent with our previous findings based on magnetotransport measurements on Sb$_2$Te$_3$ layers, in which a high-mobility two-dimensional channel was identified that corresponds to these surface state~\cite{Weyrich17}. Note that the oscillation ampitude in \ref{fig:Sb2Te3-100nm-ring-AB} is small compared to the conductance quantum. This can be due to many reasons, e.g. resistances in series from the legs and contacts as well as a suppressed amplitude due to disorder interaction which can also be seen in the simulations in \ref{fig:Conductance}. 

Based on the Shubnikov-de Haas oscillation measurements, it was found that the surface-state carrier concentration is about an order of magnitude lower than the total one. Additional information on the relevant transport regime can be found in the Supplementary Material SI.

To better understand the impact of elastic scattering due to disorder on the AB effect of the ring for bulk versus topological surface states, we perform quantum transport simulations. We make use of the quantum transport simulation package Kwant~\cite{Groth14}, the efficient parallel sparse direct solver MUMPS~\cite{Amestoy01}, and the Adaptive package~\cite{Nijholt22} to efficiently sample the parameter space, i.e., energy and flux. We employ the same tight-binding modeling approach as in Refs.~\cite{Moors18,Koelzer21} and refer to these works for more details. Disorder is considered by adding a randomly fluctuating on-site energy with a characteristic disorder strength $S_\mathrm{dis}$ in the scattering region of the tight-binding model.
In Fig.~\ref{fig:Conductance}, the simulation results are summarized. The assumed sample geometry for the simulation is depicted in Fig.~\ref{fig:Conductance}a. We compare the magnetoconductance of a bulk metallic ring with that of a bulk-insulating TI, where the bulk metallic states are described as a free electron gas with an effective mass that is appropriate for the bulk states of Sb$_2$Te$_3$ near the Fermi level. The energy window is chosen for both systems such that they have a comparable magnetoconductance in the clean limit without disorder (cf. Fig.~\ref{fig:Conductance}b). As can be seen in Fig.~\ref{fig:Conductance}c, without any disorder, the metallic ring displays the most pronounced AB effect, with the appearance of many higher harmonics (cf. Fig.~\ref{fig:Conductance}e). When disorder is introduced, however, the magnetoconductance oscillations are quickly suppressed, as well as the conductance itself. The bulk metallic states are easily driven into a highly diffusive regime by disorder, which hinders the transport along the ring and the corresponding AB signature. For the TI surface states, the behavior is quite different (cf. Figs.~\ref{fig:Conductance}d and f). While the AB peak and its harmonics in the Fourier spectrum are not so pronounced as for the metallic ring in the clean limit, disorder has a much weaker impact on the conductance and its flux quantum-periodic oscillations. This reflects the resilience of TI against elastic backscattering in the presence of disorder, which is also observed in straight nanoribbons and multi-terminal junctions~\cite{Dufouleur13,Cho15,Ziegler18,Rosenbach20,Koelzer21}. Because of their spin-momentum locking properties and being bound to the surface, a robust (quasi-)ballistic transport regime can be established for TI surface states even in the presence of relatively strong disorder throughout the ring geometry.

\subsection{Conductance fluctuations}

\begin{figure}[htb]
    \centering
    \includegraphics[width=\linewidth]{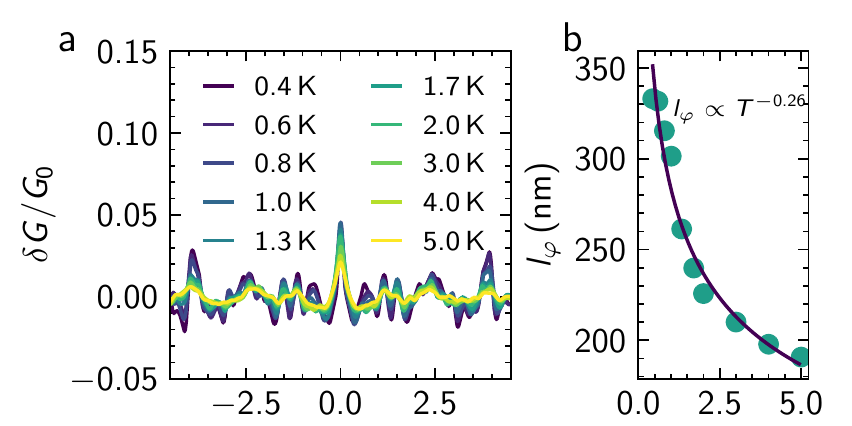}
    \caption{Universal conductance fluctuations:
    (a) Normalized fluctuation pattern present in the magnetoconductance measurements of the Sb$_2$Te$_3$ ring structure presented earlier, with $G_0=2e^2/h$. The magnetic field is oriented perpendicular to the ring structure.
    (b) Phase-coherence length $l_{\varphi}$ of the ring structure extracted from the correlation field $B_c$ as a function of temperature. The blue line indicates an exponential decrease of the phase-coherence length following $T^{-0.26}$. } 
    \label{fig:Sb2Te3-100nm-ring-UCF}
\end{figure}

In order to gain more information on the phase-coherent transport, the investigation of Aharonov-Bohm oscillations is followed up by an analysis of the conductance fluctuations also present in the magnetoconductance shown in Fig.~\ref{fig:Sb2Te3-100nm-ring-AB}a. The temperature dependence of $l_{\varphi}$ relevant for this phenomena can be determined from the correlation field $B_c$. This quantity is extracted from the normalized fluctuation patterns in the magnetoconductance $\delta G/G_0$ depicted in Fig.~\ref{fig:Sb2Te3-100nm-ring-UCF}a obtained after subtracting the slowly varying background and filtering out the Aharonov-Bohm oscillations.

It can clearly be seen that the fluctuation amplitude substantially decreases with increasing temperature, while the pattern itself is consistent over all temperatures. The correlation field $B_c$ is determined using the autocorrelation function: $F(\Delta B)=\langle \delta G(B+\Delta B)\delta G(B)\rangle$ \cite{Lee87}. Here, the full-width half maximum $F(B_c)=\frac{1}{2}F(0)$ defines $B_c$. In the diffusive regime, $l_{\varphi}$ can be determined using $l_{\varphi} \approx \gamma \phi_0/B_c d$ \cite{Beenakker88}, with $d$ the width of the ring arms and the width of leads to the ring, which is \SI{50}{nm} in our case. For the pre-factor $\gamma$ we choose $0.42$ \cite{Beenakker88} for $l_\varphi$ larger than the thermal length (see Supplementary Material SI). The resulting values of $l_\varphi$ determined from the correlation field are shown in Fig.~\ref{fig:Sb2Te3-100nm-ring-UCF}b. As indicated by the blue line the decrease of the phase-coherence length $l_{\varphi}$ with increasing temperature can be fitted by a dependency of $l_{\varphi} \propto T^{-0.26}$. The temperature dependence of the sample is slightly lower than the expected dependence of $T^{-1/3}$ for a quasi one-dimensional system \cite{Altshuler82}. The maximum of $l_{\varphi}=\SI{330}{nm}$ is smaller than the corresponding value determined from the Aharonov-Bohm oscillations. We attribute this discrepancy to different contributions to the overall phase-coherent transport. As outlined in the previous section, we concluded that the Aharonov-Bohm oscillations originate from (quasi-)ballistic transport of topologically protected surface states. In contrast, conductance fluctuations by nature only show up in the diffusive transport regime. We can therefore attribute the appearance of conductance fluctuations to diffusive bulk transport.  

\subsection{Nanoribbon measurements}

\begin{figure*}
    \centering
    \includegraphics[width=0.95\linewidth]{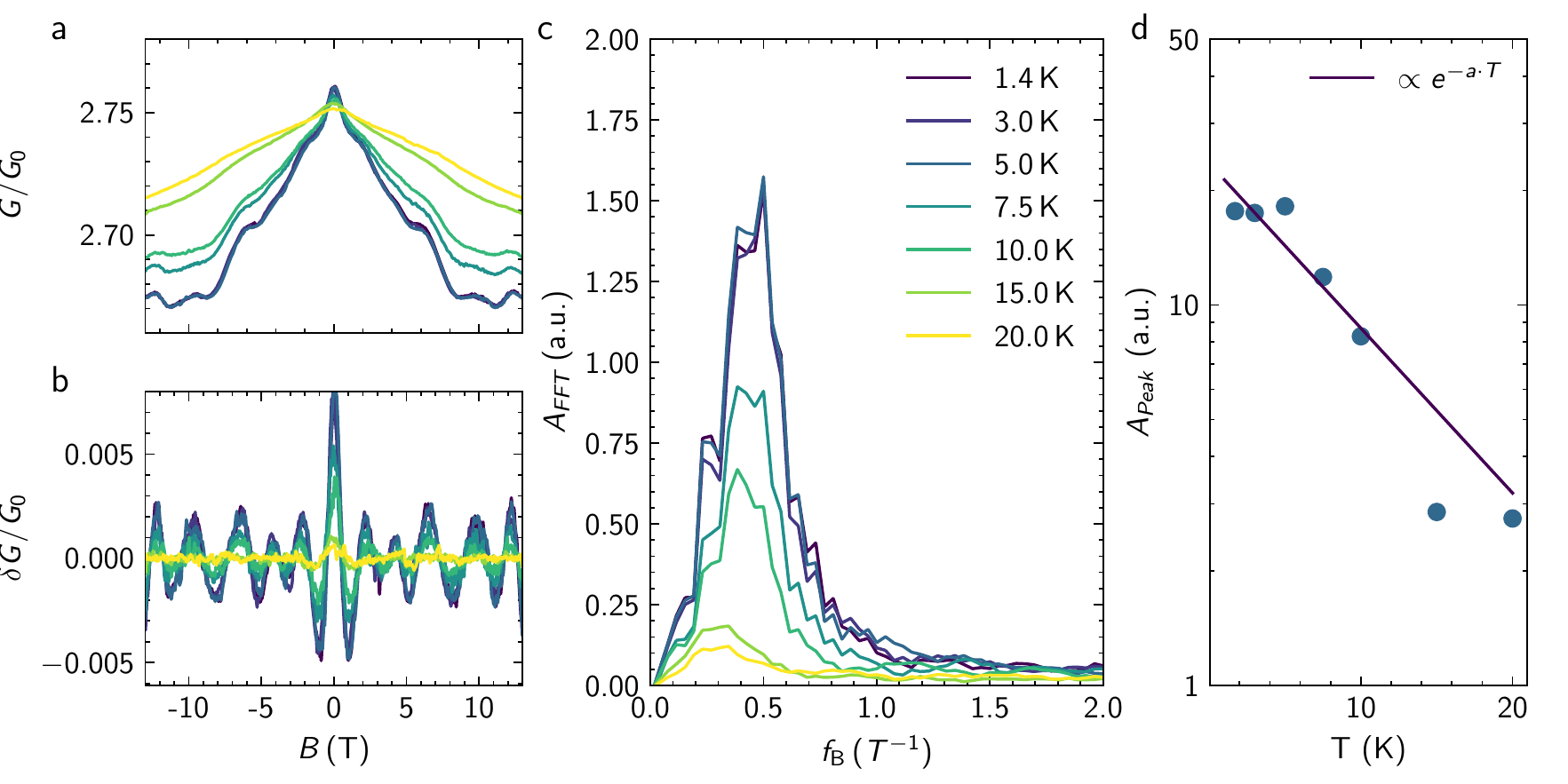}
    \caption{Magnetotransport in a nanoribbon:
    (a) Magnetoconductance of a nanoribbon structure (sample C) under application of an axial in-plane magnetic field in the range of \SI{1.4}{K} to \SI{20}{K}. The legend can be found in (c).
    (b) Corresponding magnetoconductance after subtracting the slowly varying background signal.
    (c) Fourier transform of the magnetoconductance data shown in (b) at different temperatures.
    (d) Amplitude of the peak in the FFT spectrum as a function of temperature. The solid line represents the fit.
    } 
    \label{fig:Sb2Te3_100nm-ribbon}
\end{figure*}

So far we could show that Aharonov-Bohm oscillations appear in planar ring structures. The transport regime was identified to be in the ballistic regime. However, for the topological surface states of TI nanoribbons, Aharonov-Bohm-type oscillations are also expected to show up under the application of an axial magnetic field. In this case, the oscillations originate from the interference of topologically protected surface states enclosing the magnetic flux penetrating the cross section of the nanoribbon. We verify this effect as well by measuring a 100-nm-wide Sb$_2$Te$_3$ nanoribbon from the same growth run as the planar ring under application of an in-plane field along the nanoribbon axis. Figure~\ref{fig:Sb2Te3_100nm-ribbon}a shows the corresponding magnetoconductance measurements at temperatures in the range of \SI{1.7}{K} to \SI{20}{K} under application of a magnetic field up to $\pm$\SI{13}{T}.

The data shows a peak at zero magnetic field, which can be attributed to weak antilocalization~\cite{Koelzer20,Rosenbach20}. On top of this feature, clear low-frequency magnetic field-dependent oscillations are observed. To distinguish the magnetic field-dependent oscillations from the weak antilocalization behaviour of the nanoribbon, a smooth background has been subtracted by applying a first-order Savitzky–Golay filter. The corresponding curves are depicted in Fig.~\ref{fig:Sb2Te3_100nm-ribbon}b. A clear peak at a frequency of around \SI{0.45}{T^{-1}} can be seen in the Fourier spectrum in Fig.~\ref{fig:Sb2Te3_100nm-ribbon}c. 
The cross-sectional area determined from the frequency is $1.86 \times 10^{-15}\,\mathrm{m}^2$, which is matching very well to the cross-sectional area of $2\times 10^{-15}\,\mathrm{m}^2$, which is determined from the film thickness of \SI{20}{nm} at a nanoribbon width of \SI{100}{nm}. The temperature dependence of the peak suggests a strong dependence on the phase-coherence of our carriers. The peak amplitude should vanish with temperature corresponding to a vanishing phase-coherence length of a carrier as a function of temperature. Similarly to the planar ring structures we determined the decay of the Aharonov-Bohm integrated peak height in the FFT spectrum shown in Fig.~\ref{fig:Sb2Te3_100nm-ribbon}b with temperature. Once again the decay follows an exponential dependence according to $\mathcal{A} \propto \exp(-P/l_\varphi(T))$, with $P$ the nanoribbon perimeter. The temperature dependence indicates that the transport is quasi-ballistic~\cite{Dufouleur13,Rosenbach20,Kim20}. From the fit we deduced a phase-coherence length of $l_\varphi(T = 2\mathrm{K}) = 600$\,nm at \SI{2}{K}. The observed oscillations can be attributed to phase-coherent oscillations around the perimeter of the nanoribbon. This effectively proves the existence of surface states in the investigated material.

\section{Conclusion}

From the temperature dependence of the Aharonov-Bohm oscillation amplitude Sb$_2$Te$_3$ ring interferometers we found that the phase-coherent transport takes place in the quasi-ballistic regime. Comparing the quantum transport simulations of a metallic and topological insulator ring structures, we conclude that the quasi-ballistic transport is attributable to the topologically protected surface states. The underlying reason is that these states are resilient against elastic backscattering in the presence of disorder, unlike the diffusive bulk states. In addition to the periodic Aharonov-Bohm oscillations, the magnetoconductance trace also contains irregular conductance oscillations. Since the appearance of this phenomena requires transport in the diffusive regime, we conclude that transport in the bulk channel is responsible in this case. Finally, on straight nanoribbons fabricated in the same growth run, regular Aharonov-Bohm oscillations are observed under the application of an axial magnetic field. As for the planar ring structures a quasi-ballistic regime was identified. Our investigation on planar ring structures as well as on straight nanowire thus leads us to the conclusion, that the phase-coherent transport in lateral as well as in transverse direction is quasi-ballistic. Furthermore, it seems that the transport in the topological surface states is decoupled from the phase-coherent diffusive transport in the bulk channel. The present work is an important milestone to the distinction between quantum transport in topologically protected surface states and in the bulk channel. Our results thus help to design future topological devices based on phase-coherent transport with topological insulator surface states.

\section*{Acknowledgments}
We thank Alexander Ziessen and Fabian Hassler for fruitful discussions, Herbert Kertz for technical assistance, and Florian Lentz and Stefan Trellenkamp for electron beam lithography. This work was partly funded by the Deutsche Forschungs\-gemein\-schaft (DFG, German Research Foundation) under Germany’s Excellence Strategy - Cluster of Excellence Matter and Light for Quantum Computing (ML4Q) EXC 2004/1 – 390534769. D.H.\ and K.M.\ acknowledge the financial support by the Bavarian Ministry of Economic Affairs, Regional Development and Energy within Bavaria’s High-Tech Agenda Project "Bausteine für das Quantencomputing auf Basis topologischer Materialien mit experimentellen und theoretischen Ansätzen" (grant allocation no.\ 07 02/686 58/1/21 1/22 2/23).

\bibliographystyle{apsrev4-2}

\clearpage
\widetext


\setcounter{section}{0}
\setcounter{equation}{0}
\setcounter{figure}{0}
\setcounter{table}{0}
\setcounter{page}{1}
\makeatletter
\renewcommand{\thesection}{S\Roman{section}}
\renewcommand{\thesubsection}{\Alph{subsection}}
\renewcommand{\theequation}{S\arabic{equation}}
\renewcommand{\thefigure}{S\arabic{figure}}
\renewcommand{\figurename}{Supplementary Figure}
\renewcommand{\bibnumfmt}[1]{[S#1]}
\renewcommand{\citenumfont}[1]{S#1}

\begin{center}
\textbf{Supplementary Material: Aharonov-Bohm interference and phase-coherent surface-state transport in topological insulator rings}
\end{center}

\section{Measurements on Hall bar structures}

\begin{figure*}[h!]
    \centering
    \includegraphics[width=0.35\textwidth]{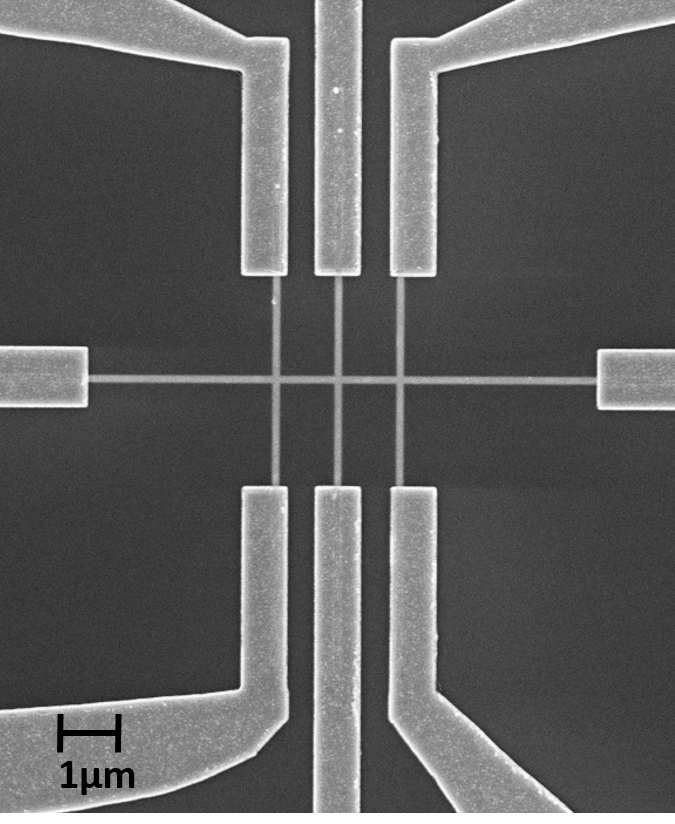}
    \caption{Scanning electron micrograph of the 500-nm-wide Hall bar structure.} 
    \label{fig:Hall-bar}
\end{figure*}

Basic Hall measurements have been conducted on a 500-nm-wide Hall bar structure in order to obtain the electronic material properties of the Sb$_2$Te$_3$ topological insulator thin film. The Hall bar was selectively grown in the same growth run as the ring and nanoribbon structures. Supplementary Figure~\ref{fig:Hall-bar} shows an electron beam micrograph of the Hall bar sample.  
The topological insulator material is grown with a thickness of about \SI{15}{nm}, which is thinner than for the ring and nanoribbon structures owing to the dependence of the growth rate on the window area in the growth mask. A standard lock-in amplifier setup and a variable temperature insert with a base temperature of \SI{1.5}{K} were used for the measurements. In Supplementary Figure~\ref{fig:Hall-bar-measurements}(a) and (b) the Hall resistance as well as the longitudinal magnetoresistance are shown. The system shows $p$-type behaviour. This is visible from the slope of the Hall resistance. The Hall resistance increases linearly indicating an effective single-channel transport. The sample exhibits a monotonic positive magnetoresistance as well as a weak antilocalization feature manifesting itself as a cusp-like dip in resistance around zero field \cite{Weyrich17}. A hole concentration of $n=$\SI{7.4}{\times 10^{13}\,cm^{-2}} and a mobility of $\mu=$\SI{152}{cm^2/Vs} were obtained from the Hall and longitudinal resistance measurements. 
\begin{figure*}[h]
    \centering
    \includegraphics[width=0.95\textwidth]{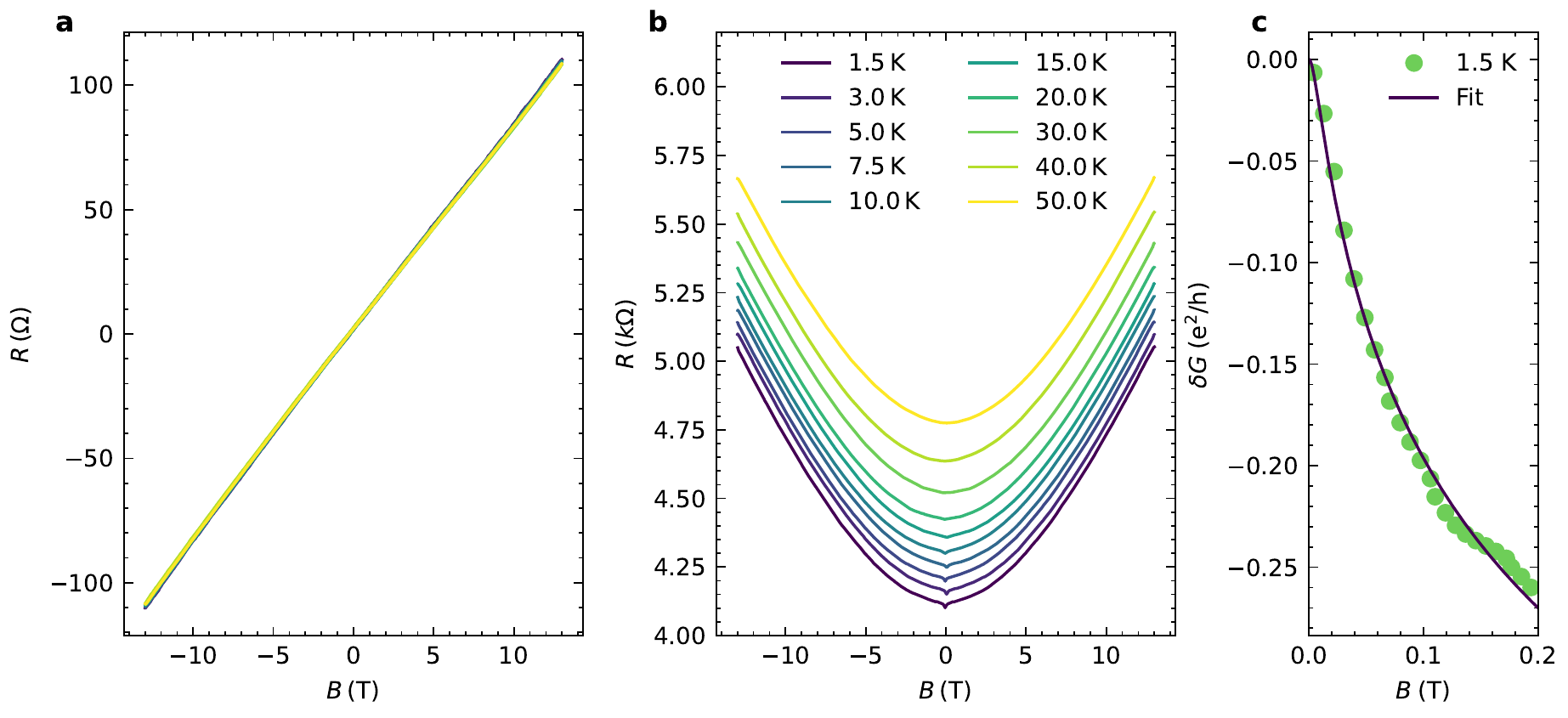}
    \caption{(a) Hall  resistance as a function of magnetic field for various temperatures of the Hall bar structure. (b) Corresponding longitudinal resistance as a function of magnetic field. (c) Measured magnetoconductance in a magnetic field range of 0 to \SI{0.2}{T} and fit according to the Hikami-Larkin-Nagaoka theory.} 
    \label{fig:Hall-bar-measurements}
\end{figure*}

Taking the relatively large carrier concentration into account, we assume that most of the carriers are located in the bulk valence band. Indeed, we estimated the bulk carrier concentration to be a factor of more than ten larger than the concentration in the surface states. The dominating bulk contribution also results in an effective single-channel transport indicated by the linear increase of the Hall resistance shown in Supplementary Figure~\ref{fig:Hall-bar-measurements}(a). Thus, in order to determine the bulk transport parameters we neglect the contribution of the holes in the surface states. The approximate effect hole mass $m^*=m_0$, with $m_e$ the free electron mass, was extracted from the band structure calculation of Zhang \textit{et al.} \cite{Zhang09}. Assuming that measured mobility is mainly governed by the bulk carriers we obtained an elastic scattering time of $\tau_e=(m^*\mu)/e$ of $4.8 \times 10^{-14}$\,s which together with an estimated Fermi velocity $v_\mathrm{F}$ of $1.3 \times 10^5$ m/s results in a diffusion constant of $\mathcal{D}=\frac{1}{3} v_\mathrm{F}^2 \tau_e \approx 0.41 \times 10^{-3}$\,m$^2$/s and an elastic mean free path $l_e=v_\mathrm{F} \tau_e \approx 11$\,nm. The bulk Fermi velocity was calculated from $v_\mathrm{F}=(\hbar/m^*) (3 \pi^2 n_\mathrm{3D})^{1/3}$, with $n_\mathrm{3D}$ the three-dimensional hole concentration. Having $\mathcal{D}$ available the thermal length defined by $l_T=\sqrt{\hbar\mathcal{D}/k_\mathrm{B}T}$ can be calculated. At a temperature of \SI{1}{K} the thermal length is approximately \SI{56}{nm}. Since the thermal length is in any case smaller than the phase-coherence length of bulk carriers determined from conductance fluctuations, we assumed a scaling factor of $\gamma= 0.42$ for the relation between the correlation field with $l_\varphi$ in the main manuscript.   

\section{Aharonov--Bohm oscillations}

We also investigate the magnetotransport of the second Aharonov-Bohm ring (sample B). For comparison, in Supplementary Figures~\ref{fig:devices}(a) and (b) electron beam micrographs are shown for sample A and B, respectively. The outer radius of sample B is \SI{200}{nm}, thus \SI{50}{nm} larger than the one of sample A. 
\begin{figure*}[]
    \centering
    \includegraphics[width=0.4\textwidth]{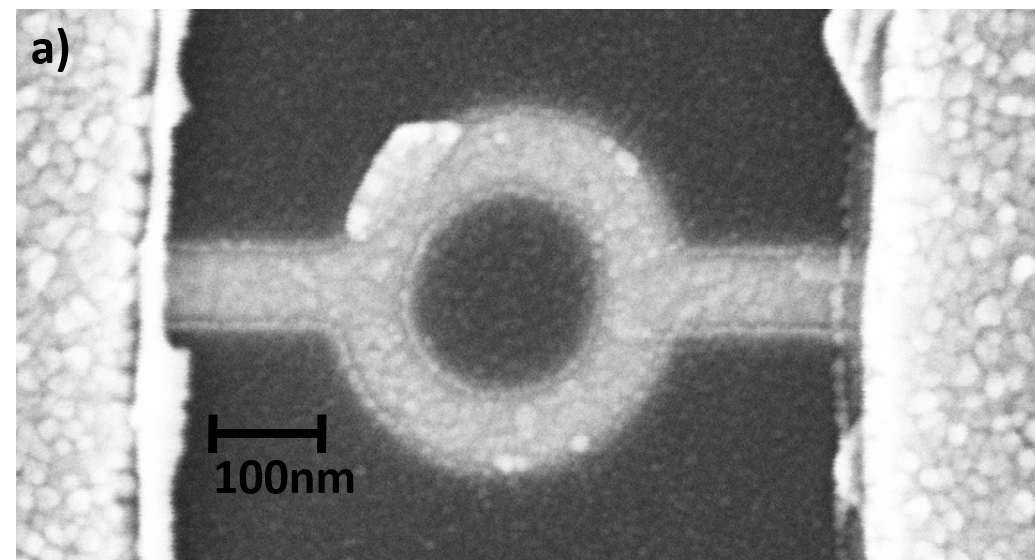}
    \includegraphics[width=0.4\textwidth]{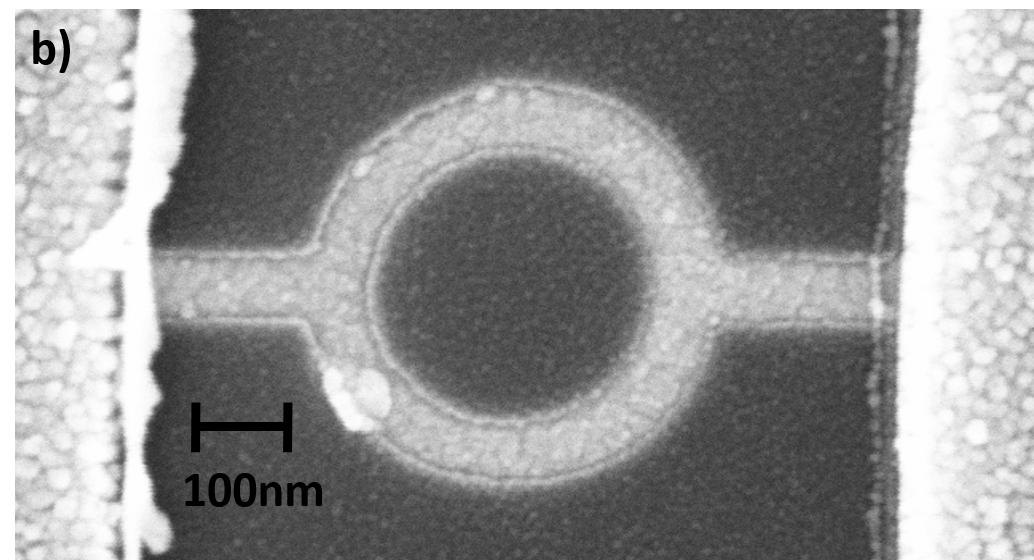}
    \caption{(a) Scanning electron micro-graph of ring sample A in comparison to sample B which is shown in (b).} 
    \label{fig:devices}
\end{figure*}

In Supplementary Figure~\ref{fig:Sb2Te3-150nm-ring-AB}(a) the normalized magnetoconductance $G/G_0$ of sample B is shown, with $G_0=2e^2/h$ and the temperatures varied between \SI{0.4}{K} to \SI{4.0}{K}.
\begin{figure*}[htb]
    \centering
    \includegraphics[width=0.95\textwidth]{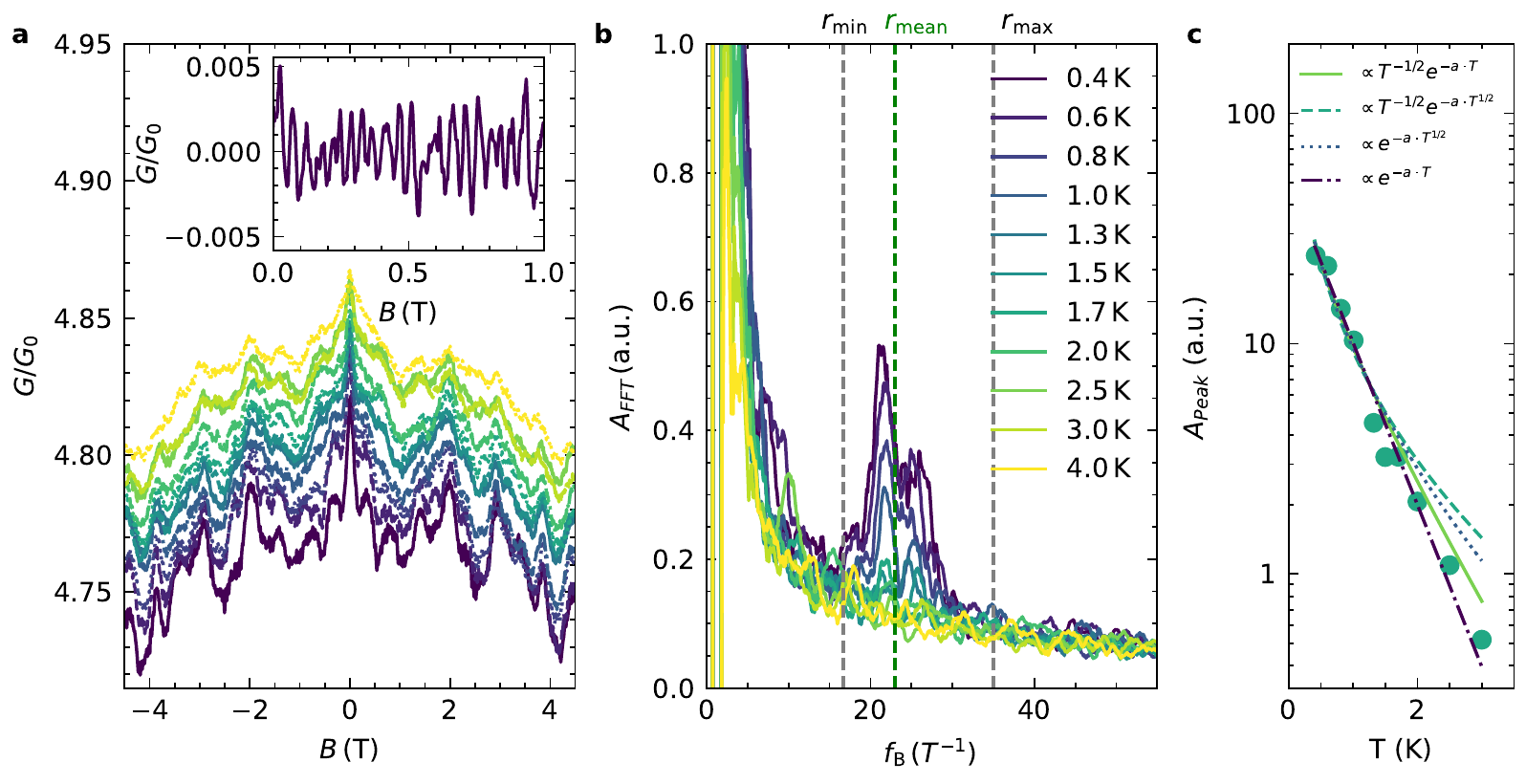}
    \caption{Magnetotransport of ring structures:
    (a) Normalized magnetoconductance of sample B at temperatures in the range of \SI{0.4}{K} to \SI{4.0}{K}, with $G_0=2e^2/h$. The inset shows a detail of the oscillations at \SI{400}{mK} in a smaller magnetic field range. The measured period $\Delta B$ of \SI{47}{mT}, fits well to the peak observed in the FFT of the measured data. The legend for the curves belonging to different temperatures is given in (b).
    (b) Fourier spectrum of the magnetoconductance shown in (a). The frequency spectrum was smoothed with a moving average. The expected frequency according to the minimum and maximum radius as well as the mean radius are indicated by dashed lines.
    (c) Integrated amplitude $\mathcal{A}$ of the peak in the FFT at \SI{23}{T^{-1}} as a function of temperature. The solid violet line represents an exponential fit with $\exp(-aT)$ corresponding to ballistic transport without thermal broadening. Also shown are fits assuming diffusive transport, i.e. with $\exp(-aT^{1/2})$, as well as thermal broadening, i.e. with prefactor $\sqrt{T}$.} 
    \label{fig:Sb2Te3-150nm-ring-AB}
\end{figure*}
The magnetoconductance show the same kind of features as for sample A, i.e., a peak at zero resistance due to the weak antilocalization effect as well as conductance fluctuations. Superimposed, one also finds regular Aharonov-Bohm oscillations. This becomes clear by having a look on the magnification of a smaller magnetic field region depicted in Supplementary Figure~\ref{fig:Sb2Te3-150nm-ring-AB}(a) (inset). From that we extracted an oscillation period of $\Delta B=\SI{47}{mT}$. The period $\Delta B$ fits very well to oscillations with a flux period of $\phi_0=\Delta B \cdot A$, with $\phi_0=h/e$ the magnetic flux quantum and $A=\pi r_\mathrm{mean}^2$ the ring area for the mean radius $r_\mathrm{mean}$ of \SI{175}{nm}.

As for ring sample A, the periodic features in the magnetoconductance are analyzed by a fast Fourier transform (FFT), as shown in Supplementary Figure~\ref{fig:Sb2Te3-150nm-ring-AB}(b). In the FFT spectrum we resolve a distinct peak at a frequency $f_B$ of about \SI{21.5}{T^{-1}}, which fits to the expected value for the mean radius of the ring indicated by the vertical dashed green line in Supplementary Figure~\ref{fig:Sb2Te3-150nm-ring-AB}(b). Increasing the temperature causes a decrease of the peak height. In contrast to sample A no second harmonic feature is observed in the Fourier spectrum.

The phase-coherence length $l_\varphi$ was determined from the decrease of the integrated peak height $\mathcal{A}(T)$ at \SI{21.5}{T^{-1}} in the Fourier spectrum with temperature. The integration is performed between frequencies corresponding to the minimum and maximum ring radii (cf. Supplementary Figure~\ref{fig:Sb2Te3-150nm-ring-AB}(b)). As for sample A, we assumed an exponential decay $\mathcal{A} \propto \exp(-\pi r_\mathrm{mean}/l_\varphi(T))$, with $\pi r_\mathrm{mean}= \SI{550}{nm}$ the length of one of the ring arms. The peak height decay is very well fitted by an exponential decrease with $l_\varphi(T) \sim T^{-1}$, as shown in Supplementary Figure~\ref{fig:Sb2Te3-150nm-ring-AB}(c). From the fit, we obtain a phase coherence length of $l_\varphi = \SI{859}{nm}$ at a temperature of \SI{0.4}{K}.

Supplementary Figure~\ref{fig:Sb2Te3-150nm-ring-AB}(c) also shows fits for the peak amplitude for different transport regimes. Generally the temperature dependence of the Aharonov-Bohm  oscillation amplitude can be described by \cite{Milliken87,Washburn86}     
\begin{equation}
    \delta G \propto \left( \frac{E_\mathrm{Th}}{k_\mathrm{B}T}\right)^{(1/2)} \exp(-\pi r /l_\varphi) \, ,  
\end{equation}
where $E_\mathrm{Th}$ is the Thouless energy \cite{Edwards72}, i.e., for diffusive conductors it is given by $E_\mathrm{Th}=\hbar \mathcal{D}/L^2$, with $\mathcal{D}$ the diffusion constant and $L$ the ring circumference. For the ballistic case the situation for the Thouless energy is more subtle~\cite{Altland96}. In case that the thermal energy $k_\mathrm{B}T$ is larger than $E_\mathrm{Th}$ thermal broadening occurs, leading to a pre-factor $T^{-1/2}$ of the oscillation amplitude. In case that $E_\mathrm{Th}$ is larger than $k_\mathrm{B}T$ this factor can be neglected. Furthermore, the temperature dependence of $l_\varphi$ differs for the diffusive and ballistic case. Ludwig and Mirlin theoretically found that for the diffusive case the phase-coherence length is proportional to $T^{-1/2}$~\cite{Ludwig04}, whereas in the ballistic case one expects $l_\varphi \sim T^{-1}$ \cite{Seelig01}. In diffusive metallic and semiconducting rings a temperature dependence of $l_\varphi \sim T^{-p}$ with $p$ in between $0.5$ and $0.75$ was observed~\cite{Milliken87,Appenzeller95}. Whereas in clean semiconductor heterostructure rings \cite{Hansen01,Lin10} as well as in clean graphene ring structures~\cite{Dauber17} a temperature dependence of $l_\varphi \sim T^{-1}$ was found, indicating a ballistic transport regime~\cite{Seelig01}. In topological insulator nanoribbons exposed to a magnetic field along the ribbon axis, Aharonov-Bohm oscillations are observed due to the presence of tubular topologically protected surface states. These systems were also found to be in the ballistic regime \cite{Dufouleur13,Ziegler18,Rosenbach20}. In Supplementary Figure~\ref{fig:Sb2Te3-150nm-ring-AB}(c), fits assuming different scenarios are shown, i.e., thermal broadening vs. no thermal broadening and diffusive vs. ballistic transport. Obviously, a good fit was only obtained for the ballistic case without thermal broadening implying a large Thouless energy and a long mean free path. The latter can be explained by the reduced backscattering of carriers in topologically protected surface states. 

\section{Weak antilocalization analysis}

As mentioned in the previous section, the magnetoresistance shown in Supplementary Figures~\ref{fig:Hall-bar-measurements}(b) exhibits a small cusp-like dip at zero field which can be assigned to the weak antilocalization effect. The weak antilocalization effect for two-dimensional systems can be described by the the Hikami--Larkin--Nagaoka (HLN) formula  \cite{Hikami80}. A fitting to the experimental magnetoconductance (cf. Supplementary Figure~\ref{fig:Hall-bar-measurements}(c)) results in a phase-coherence length of \SI{225}{nm} with an $\alpha$-factor of $-0.37$ at base temperature. The $\alpha$ factor is a measure of the number of transport channels and is close to the value of $-0.5$ corresponding to a single channel. Our values of $\l_\varphi$ and $\alpha$ indicate a strong bulk conduction of the system, which is typical for this type of samples as the charge carrier concentration is significantly higher and the mobility is lower than expected for Dirac surface states. The phase-coherence length extracted here fits well to the value obtained from the conductance fluctuations, indicating that both effects are governed by bulk carriers.  

\bibliographystyle{apsrev4-2}

\end{document}